\documentclass[nofootinbib,noeprint,aps,prl,reprint,superscriptaddress]{revtex4-1}
\usepackage[T1]{fontenc} % Use modern font encodings
\usepackage{color}
\usepackage{textcomp}
\usepackage{siunitx}
\usepackage{xspace}
\usepackage{mathtools}
\usepackage{mathpazo}
\DeclarePairedDelimiter\abs{\lvert}{\rvert}%
\DeclarePairedDelimiter\norm{\lVert}{\rVert}%

\makeatletter
\let\oldabs\abs
\def\abs{\@ifstar{\oldabs}{\oldabs*}}
\let\oldnorm\norm
\def\norm{\@ifstar{\oldnorm}{\oldnorm*}}
\makeatother

\newcommand{\ie}{i.\,e.}

\newcommand{\sio}{SiO$_2$\xspace}
\newcommand{\is}{I_\text{S}}
\newcommand{\vsd}{V_\text{SD}}
\newcommand{\vbg}{V_\text{BG}}
\newcommand{\ic}{I_\text{SW}}
\newcommand{\ir}{I_\text{R}}

\newcommand{\gn}{G_\text{N}}

\newcommand{\icrn}{I_\text{SW}R_\text{N}}

\newcommand{\dvdi}{\partial{V_\text{SD}}/\partial{I_\text{S}}}
\newcommand{\didv}{\partial{I_\text{S}}/\partial{V_\text{SD}}}
\newcommand{\didvb}{\partial{I_\text{D}}/\partial{V_\text{SD}}}
\newcommand{\vrms}{V_\text{rms}}
\newcommand{\tc}{T_\text{C}}
\newcommand{\tcal}{T_\text{C,Al}}
\begin{document}
%\preprint{}

\title{Multiple Andreev reflections and Shapiro steps in a Ge-Si nanowire Josephson junction}

% repeat the \author .0613895475. \affiliation  etc. as needed


\author{Joost Ridderbos}
\affiliation{MESA+ Institute for Nanotechnology, University of Twente, P.O. Box 217, 7500 AE Enschede, The Netherlands}
\author{Matthias Brauns}
\affiliation{MESA+ Institute for Nanotechnology, University of Twente, P.O. Box 217, 7500 AE Enschede, The Netherlands}

%\author{Jie Shen}
%\affiliation{QuTech and Kavli Institute of Nanoscience, Delft University of Technology, 2600 GA Delft, The Netherlands}
%\author{Folkert K. de Vries}
%\affiliation{QuTech and Kavli Institute of Nanoscience, Delft University of Technology, 2600 GA Delft, The Netherlands}

\author{Ang Li}
\affiliation{Department of Applied Physics, Eindhoven University of Technology, P.O. Box 513, 5600 MB Eindhoven, The Netherlands}
%\alsoaffiliation{Beijing Key Laboratory of Microstructure and Property of Advanced Materials, Beijing University of Technology, Beijing, China}
\author{Erik P. A. M. Bakkers}
\affiliation{Department of Applied Physics, Eindhoven University of Technology, P.O. Box 513, 5600 MB Eindhoven, The Netherlands}
\affiliation{QuTech and Kavli Institute of Nanoscience, Delft University of Technology, 2600 GA Delft, The Netherlands}
\author{Alexander Brinkman}
\affiliation{MESA+ Institute for Nanotechnology, University of Twente, P.O. Box 217, 7500 AE Enschede, The Netherlands}
\author{Wilfred G. van der Wiel}
\affiliation{MESA+ Institute for Nanotechnology, University of Twente, P.O. Box 217, 7500 AE Enschede, The Netherlands}
\author{Floris A. Zwanenburg}
%\email[Corresponding author, e-mail: ]{f.a.zwanenburg@utwente.nl}
\email{f.a.zwanenburg@utwente.nl}
\affiliation{MESA+ Institute for Nanotechnology, University of Twente, P.O. Box 217, 7500 AE Enschede, The Netherlands}


%\date{\today}
%\begin{document}

\begin{abstract}
We present a Josephson junction based on a Ge-Si core-shell nanowire with transparent superconducting Al contacts, \textcolor{black}{a building block which could be of considerable interest for investigating Majorana bound states, superconducting qubits and Andreev (spin) qubits.} We demonstrate the dc Josephson effect in the form of a finite supercurrent through the junction, and establish the ac Josephson effect by showing up to 23 Shapiro steps. We observe multiple Andreev reflections up to the sixth order, indicating that charges can scatter elastically many times inside our junction, and that our interfaces between superconductor and semiconductor are transparent and have low disorder.
\end{abstract}

%Todo:
%- Re-read whole paper (especially introduction)
%V In introduction: add ASQ and AQ refs proposal + hyperfine ref 
%V Fix fig Ic(T)
%- Add Thouless
%- Add explanation of Ic(T) model
%- Write about transparencies, one model claims 0.5, other claims 0.8, MAR looks 0.7-0.8sh for single mode.

%\pacs{}
%\keywords{}

\maketitle

\section{Introduction}

\textcolor{black}{Josephson junctions are defined as a weak link between two superconducting reservoirs, which allows a supercurrent to be transported through intrinsically non-superconducting materials, as long as the junction is shorter than the coherence length~\cite{Josephson1962,Tinkham}.} 
While early Josephson junctions realized a weak link by using thin layers of oxide, micro-constrictions, point contacts or grain boundaries~\cite{Shapiro1963,States1967,Likharev1979,Beenakker1992a,Grossman1994}, access to complex mesoscopic semiconducting materials have led to Josephson junctions in which control over the charge carrier density enables \textit{in situ} tuning of the junction transparency and critical current~\cite{Krive2004,Jarillo-Herrero2006,Katsaros2010,Mizuno2013b,Saldana2018,Hendrickx2018}. 
\paragraph{}
\textcolor{black}{Devices employing semiconducting nanowires have consequently explored a wide range of applications in a variety of material systems such as SQUIDs (superconducting quantum interference devices)~\cite{Kim2016a,Cleuziou2006a,Vigneau2019}, $\pi$-junctions~\cite{Cleuziou2006a,VanDam2006b,Jorgensen2007a,Delagrange2018}, % CNT, InAs, CNT, PbS, CNT
and Cooper pair splitters~\cite{Hofstetter2009,Tan2015a}. Additionally, superconducting trans- and gatemon qubits have been successfully realized using InAs nanowires~\cite{DeLange2015,Larsen2015} and carbon nanotubes~\cite{Mergenthaler2019}, while considering 2-dimensional materials, graphene~\cite{Kroll2018} and InAs-InGaAs quantum wells~\cite{Casparis2018} have been used.
\paragraph{}
Another field where induced superconductivity in mesoscopic junctions is key, is the undergoing experimental confirmation of Majorana fermions. This has resulted in a great number of works~\cite{Kitaev2000,Mourik2012b,Deng2016,Aguado2017,Lutchyn2018,Gul2018}, but results have been limited to only a handful of material systems.
}

%Transmon microwave quantun circuits: Delange2015,
%SC qubits: Casparis2018 (Gatemon 2DEG, MW circuit), Larsen2015 (NW), Mergenthaler2019(CNTs), Kroll2018(graph transmon)
%V Andreev bound states: Hays2018, Tosi2019 (Saclet Pothier paper), Janvier2015

%V Pi junction: Saldana2018, Delagrange2018
%V Majoranas: Kitaev2001, Lutchyn2018, Aguado2017, Gül2018, Deng2016 
%X Majorana HgTe: Wiedenmann2015a, Bocquillon2106
%V Ge 2DEG: Vigneau2019, Hendrickx2018,
%Review SC qubit in general: Wendin2017

\paragraph{}
\textcolor{black}{In this work we present a Josephson junction with transparent high-quality interfaces based on semiconducting Ge-Si core-shell nanowires. This material system has proven itself in the realm of normal-state quantum dots~\cite{Hu2007,Hu2012,Higginbotham2014,Higginbotham2014a,Brauns2015,Brauns2015b,Brauns2016,Zarassi2017a,Froning2018}, but apart from a limited number of reports~\cite{Xiang2006,Su2016,Ridderbos2017,DeVries2018}, topics related to induced superconductivity are relatively unexplored.
}  
%Next to the possibility for Ge-Si nanowires to be implemented in trans- or gatemon qubits, the high presented interface quality combined with the potentially zero hyperfine interaction and strong spin-orbit interaction %addcite
%makes Ge-Si nanowires also an excellent system for the realisation of Andreev (spin) qubits with only %ASQ and AQ refs proposal 
%a limited number of experimental results existing on this topic~\cite{Janvier2015,VanWoerkom2017b,Hays2018,Tosi2019}
%Additionally, holes in Ge-Si nanowires possess the necessary ingredients to host Majorana fermions~\cite{Maier2014}: they are predicted to have strong, tunable spin-orbit coupling~\cite{Kloeffel2011,Higginbotham2014b} and have a Land\'{e} \emph{g}-factor that is tunable with electric field~\cite{Brauns2015}.

\textcolor{black}{Apart from the possibility for Ge-Si nanowires to be implemented in trans- or gatemon qubits, holes in this system possess several interesting physical properties which makes them highly suitable for hosting Majorana fermions~\cite{Maier2014} and Andreev (spin) qubits~\cite{Chtchelkatchev2003,Zazunov2003,Padurariu2010,Janvier2015,VanWoerkom2017b,Hays2018,Tosi2019}. They are predicted to have strong, tunable spin-orbit coupling~\cite{Kloeffel2011,Higginbotham2014b}, have a Land\'{e} \emph{g}-factor that is tunable with electric field~\cite{Brauns2015} and have potentially zero hyperfine interaction~\cite{Itoh2003}. The realization of a Josephson junction with transparent high-quality interfaces is a crucial step towards all the described applications for this system.
}

\paragraph{}
Using superconducting Al contacts on the Ge-Si nanowire, we will present the experimental observation of the dc Josephson effect: a finite switching current $\ic$ through the nanowire Josephson junction. We will also look at multiple Andreev reflections (MAR)~\cite{Kuhlmann1994a,Flensberg1989a} and analyze the position of the resulting conductance peaks inside the superconducting gap of Al, $\Delta_\text{Al}$. Additionally, we look at the temperature dependence of MARs and $\ic$ and finally, we irradiate our junction with microwaves resulting in Shapiro steps, the first report on the ac Josephson effect in this system. The observation of both the dc and the ac Josephson effect confirms we have a true Josephson junction.
%, which opens up the possibility for exploring more advanced systems such as SQUIDs~\cite{Cleuziou2006a} or quantum dots coupled to superconducting contacts~\cite{DeFranceschi2010,VanDam2006b,Ridderbos2017}.
%InSb MAR+Josephson:
%\cite{Nilsson2012}

\begin{figure*}
  \includegraphics{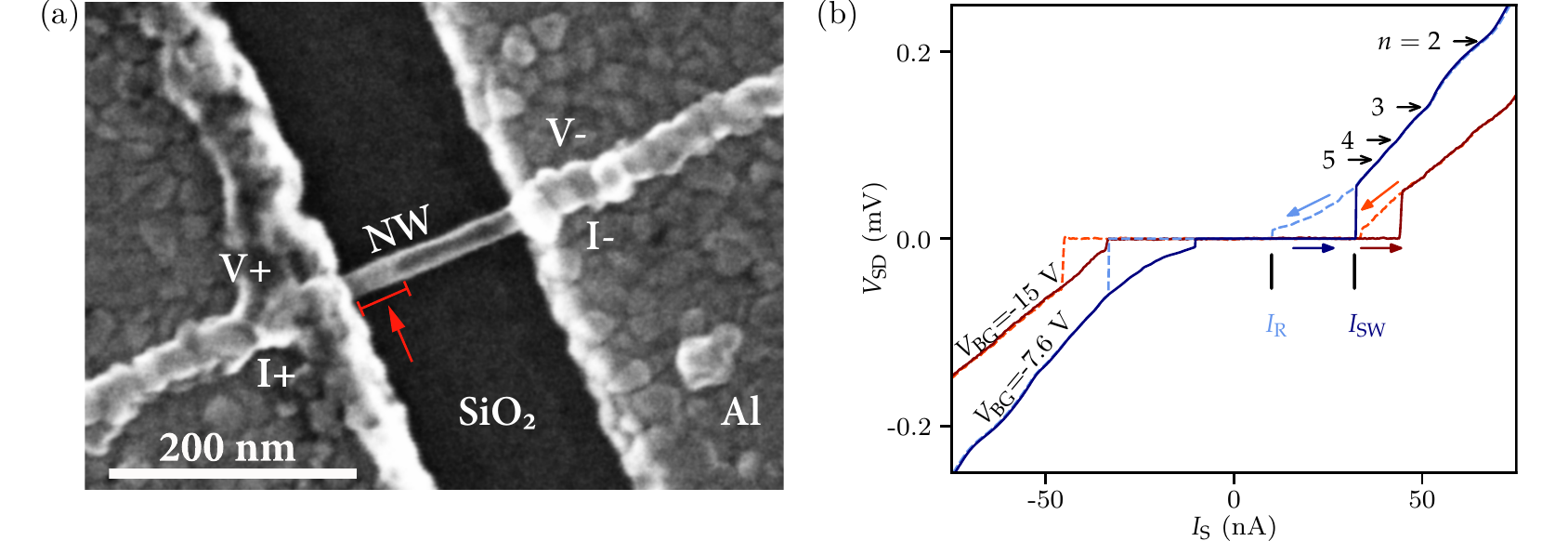}
  \caption{\textbf{dc Josephson effect in a Ge-Si nanowire.} a) False-color SEM image of the device under investigation. A nanowire with a $20$~nm diameter lies on
   the \sio covered substrate and is contacted by an Al source and drain. The channel length is \textasciitilde$150$~nm with a semiconducting island (red arrow) of \textasciitilde$50$~nm. b) $\vsd$ vs $\is$ for $\vbg=-7.6$~V and $\vbg=-15$~V. $\is$ is swept from left to right (solid) and successively from right to left (dashed) denoted by the colored arrows. $\ic$ and $\ir$ are indicated for $\vbg=-7.6$~V. Horizontal black arrows indicate `wiggles' in the curve corresponding to MAR of the $n^\text{th}$ order.} 
  \label{fig1}
\end{figure*}

\section{Nanowire Josephson junction}
Figure~\ref{fig1}a shows a SEM (scanning electron microscopy) image of the device with a channel length of \textasciitilde$150$~nm, designed for 4-terminal measurements. As described in detail in Ref.~\cite{Ridderbos2017a}, 
\textcolor{black}{Ge and/or Si inter diffuses with the Al contacts, during thermal annealing. This leaves a semiconducting island of \textasciitilde$50$~nm, which can be identified by a difference in contrast in the nanowire core on the SEM image. This has been confirmed by a TEM (transmission electron microscopy) study with an EDX (energy-dispersive x-ray) spectrum analysis on the same device (see Ref.~\cite{Ridderbos2017a}).}
\paragraph{}
We plot the sourced current $\is$ versus the measured voltage between source and drain $\vsd$ in Fig.~\ref{fig1}b. Sweeping $\is$ forward, \ie, from $0$ to finite bias, we find that the junction switches from the superconducting state to a dissipative state at a switching current of $\ic=44$~nA at a backgate voltage $\vbg=-15$ while $\ic=32$~nA at $\vbg=-7.6$~V. When sweeping backwards, \ie, from finite $\is$ to $0$, the junction returns to its superconducting state at the retrapping current $\ir$, resulting in hysteretic behavior. For a backgate voltage $\vbg=-15$~V, we find $\ir=34$~nA and a ratio $\ic/\ir=1.3$, while for $\vbg=-7.6$~V, $\ir=10$~nA and a ratio $\ic/\ir=3.4$. This indicates that our junction is underdamped~\cite{Stewart1968a} and that $\ic$, as well as the damping, depend on $\vbg$, mainly due to the changing number of subbands participating in transport and their position relative to the Fermi energy of the Al contacts. 
\textcolor{black}{As described in extensive detail in Ref.~\cite{Ridderbos2017}, the device is tunable from full depletion (with $\ic=0$) to highly transparent where $\ic>40$~nA on which this work is focused.}

\begin{figure}
  \includegraphics{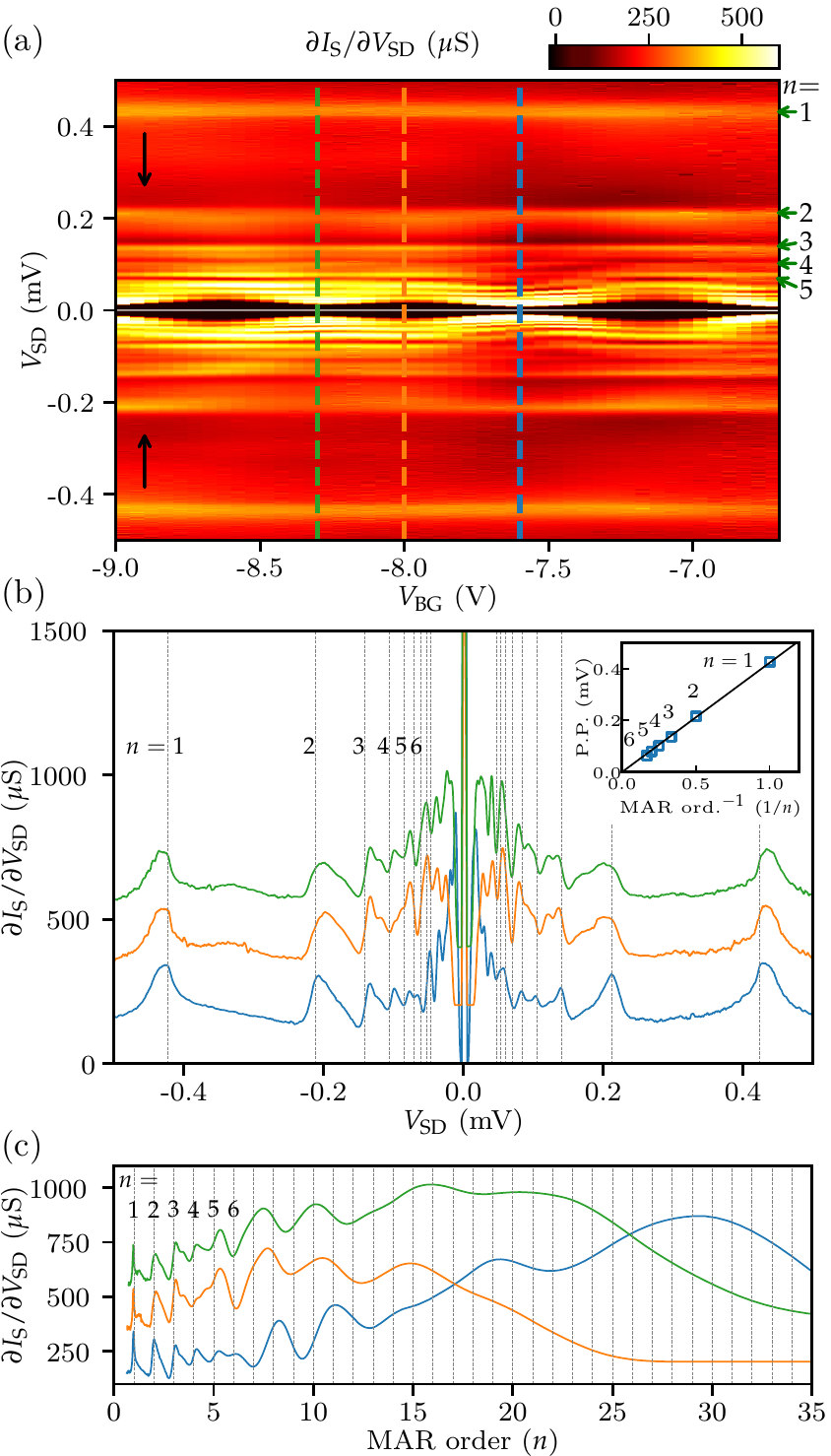}
  \caption{\textbf{Multiple Andreev reflections up to the 6$^\text{th}$ order.} a) Differential conductance $\didv$ versus $\vsd$ and $\vbg$. The black arrows indicate the sweep direction (see.~\ref{mardata}). Horizontal equipotential lines of increased conductance indicated by the green arrows correspond to MAR. Current biased measurement where the $\is$ and $\vsd$ axes were inverted (see Methods) before numerical derivation. Only return current was measured (see black arrows) in a highly hysteretic regime (see blue curve in Fig.~\ref{fig1}b) to reach the low-voltage regime. b) Single traces of $\didv$ versus $\vsd$ for three values of $\vbg$. Green, orange and blue (traces offset by $200$~$\mu$S) taken at $\vbg=\text{-}8.3$, $\text{-}8$ and $\text{-}7.6$~V (see Fig.~\ref{fig2}a, dashed lines). Vertical grey dashed lines denote expected MAR peak positions calculated by $n=2\Delta_\text{Al}/eV_\text{SD}$ for $n=1-10$. Inset: MAR peak positions (P.P.) vs inverse MAR order $1/n$ at $\vbg=-7.6$ (blue trace) for positive bias. The black line is a linear fit through zero. c) Same data as b) plotted versus $n$, only positive $\vsd$ is shown. The vertical dashed gray lines show integers of $n$ which can be matched with the MAR peaks up to $n=6$.} 
  \label{fig2}
\end{figure}

\section{Junction characteristics}
We will now establish whether our \textcolor{black}{nanowire is ballistic or diffusive}. In the ballistic case, particles traversing the junction do not scatter, except on the interfaces. In the diffusive case, particles encounter scattering sites inside the junction which leads, for example, to suppression of $\ic$~\cite{Du2008}. For a ballistic nanowire and completely transparent interfaces, one expects the normal-state conductance to appear in multiples of the conductance quantum $G_0=2\text{e}^2/h$, and the critical current in multiples of the maximum critical current for a single subband $I_\text{C,MAX}=e\Delta_\text{Al}/\hbar=51$~nA~\cite{Tinkham}. In our case, the finite interface transparency~\cite{Ridderbos2017} leads to lower observed values of both the conductance $G$ and the switching current $\ic$, where $\ic$ is suppressed by additional mechanisms such as electromagnetic coupling with the environment~\cite{Jarillo-Herrero2006}, premature switching and heating effects~\cite{Xiang2006b,Tinkham}. From experiments it is therefore not trivial to conclude whether our nanowire is diffusive or ballistic and we therefore make a quantitative estimation based on calculations.
\paragraph{}
We start with estimating the elastic scattering length using $l_\text{e}=\mu m^\star v_\text{F}/\text{e}$ \cite{Davies} with $\mu$ the hole mobility, $m^\star$ the effective hole mass and $v_\text{F}$ the Fermi velocity. We use $\mu\approx\num{3500}$~cm$^2$/Vs (determined at 4 K, see \cite{Conesa-Boj2017}) and $m^\star\approx0.5m_\text{e}$ for the mixed heavy and light holes~\cite{Kloeffel2011,Maier2014a} with $m_\text{e}$ the free electron mass. To obtain the Fermi velocity we use the solutions of the Schr\"odinger equation for a cylindrical potential well and find the expression for the Fermi energy of the $n$\textsuperscript{th} subband with quantum number $l$ as $E_{n,l}=\hbar^2\alpha_{n-1,l}^2/2m^\star R^2$~\cite{Griffiths} with $\alpha_{n,l}$ the $l$\textsuperscript{th} root of the $n$\textsuperscript{th} order Bessel function and $R$ the wire radius. For the first subband this gives $E_{1,1}\approx15$~meV, corresponding to a Fermi velocity $v_\text{F}\approx1\cdot 10^{5}$~m/s and an estimated elastic scattering length of $l_\text{e}\approx 100$~nm. 
\textcolor{black}{Using a gate lever arm $\alpha=0.02$ and the fact that the nanowire is depleted at $\vbg\approx5$~V~\cite{Ridderbos2017}, we find that in the regime $\vbg=[-7.6, -15]$~V we operate at 6 ($E_{6,1}\approx256$~meV) to 8 ($E_{8,1}\approx388$~meV) subbands, increasing $l_\text{e}$ to $\sim 400$~nm.}
\paragraph{}
As can be seen in Fig.~\ref{fig1}a, our nanowire channel length is \textasciitilde$150$~nm, but as discussed before, our semiconducting island is \textasciitilde$50$~nm. We are therefore far away from the diffusive limit $l_\text{e}<<L$ with a corresponding coherence length of $\xi_\mathrm{diff}=\sqrt{\hbar D/\pi\Delta_\text{Al}}\approx390$~nm with $D=v_\text{F}l_\text{e}$. We approach the ballistic limit $l_\text{e}>>L$, with a coherence length of $\xi_\mathrm{ball}=\hbar v_\text{F}/\pi\Delta_\text{Al}\approx380$~nm~\cite{Tinkham} independent of $l_\text{e}$. \textcolor{black}{This places the nanowire well within the ballistic limit, as is reaffirmed by the fact that the semiconducting island can be host to a single few-hole quantum dot~\cite{Ridderbos2017} and that highly-tunable normal-state devices can be host to dots of length $l>400$~nm~\cite{Brauns2015b}. By increasing the Al-nanowire interface transparencies, fully ballistic junctions could therefore be realised with lengths up to a few hundred nanometer.}

\paragraph{}
\textcolor{black}{In Ref.~\cite{Ridderbos2017} we extract an averaged $\langle\icrn\rangle=217$~$\mu$eV~$\sim\Delta_\mathrm{Al}$, close to the theoretical maximum. Since our Ge-Si segment is ballistic, the Thouless energy has only meaning in terms of the time of flight through the junction $\tau=L/v_\mathrm{F}\approx125$~fs so that $E_\mathrm{Th,ball}=\hbar / \tau\approx5.5$~mV~$\gg\Delta_\mathrm{Al}$ for the sixth subband and the induced gap is therefore $\sim\Delta_\mathrm{Al}$.}
\paragraph{}
Out of the 7 devices exhibiting superconducting transport, 3 devices showed gate-tunability and Shapiro steps. There are strong indications that in the other 4 devices the Al inter-diffusion has progressed throughout the channel (see Ref.~\cite{Ridderbos2017a}), resulting in a completely metallic superconducting device.

\section{Multiple Andreev reflections}
\label{sec:MAR}
\begin{figure*}
  \includegraphics{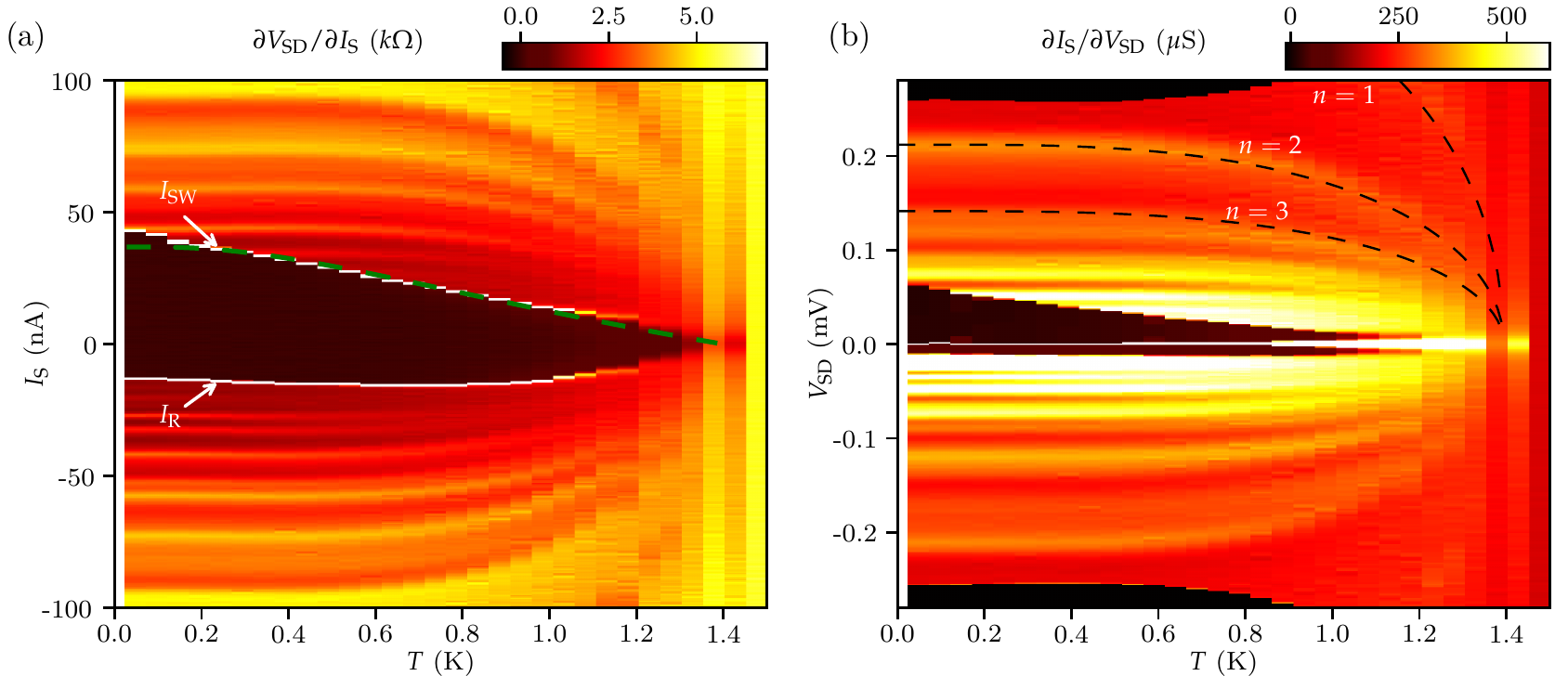}
  \caption{\textbf{Temperature dependence of MAR and $\ic$.} a) Differential resistance $\dvdi$ vs $\is$ and $T$ and b) differential conductance $\didv$ vs $\vsd$ and $T$, both for the same data with $\vbg=-13.35~$~V. $\is$ and $\vsd$ are swept from negative to positive bias. In a) $\ic$ and $\ir$ are denoted by the white arrows and the green dashed line is a fit based on the Eilenberger equations~\cite{Galaktionov2001}. Black dashed curves in b) are fits to \textcolor{black}{Eq.~\ref{bcs}.}}
  \label{fig4}
\end{figure*}

Small wiggles are visible in the $\vsd$ versus $\is$ curve for $\vbg=-7.6$~V in Fig.~\ref{fig1}b, which are a signature of MAR. This becomes clearer in the differential conductance $\didv$ for a range of $\vbg$ in Fig.~\ref{fig2}a: the wiggles in $\vsd$ translate to conductance peaks seen at values of $\vsd$ corresponding to $e\vsd(n)=2\Delta_\text{Al}/n$~\cite{Tinkham} with $n$ an integer denoting the MAR order. $n=1\text{-}5$ are indicated by the green arrows in Fig.~\ref{fig2}a for positive bias [orders $n=2\text{-}5$ are indicated in Fig.~\ref{fig1}(b)]. The strong conductance peak at $\vsd=0$ corresponds to a supercurrent and is a direct consequence of the inversion of the $\is$ and $\vsd$ axes (see Methods), which maps $\is$ onto the corresponding value of $\vsd$. Since a supercurrent implies $\vsd=0$ for a range of $\is$, this results in a strong peak in $\didvb$. The height of the oscillating black regions for $\vsd<0.05$~mV as a function of $\vbg$ is a measure for the magnitude of $\ir$ (see Methods) where the oscillations correspond to varying occupation of the subbands of a weak confinement potential in the wire~\cite{Ridderbos2017}.
\paragraph{}
The MAR conductance peaks can be more clearly distinguished in individual linetraces in Fig.~\ref{fig2}b and we focus on the blue trace at $\vbg=-7.6$ V. The finite width of the MAR peaks reflects the distribution of the DOS peak at $e\vsd=2\Delta_\text{Al}$ and is additionally broadened by phase decoherence and inelastic processes when quasiparticles traverse the channel~\cite{Nilsson2012,Tinkham}. We extract the peak positions (P.P.) in $\vsd$ of the first 6 orders and plot them versus the inverse MAR order in the inset in Fig.~\ref{fig2}b. We expect the second order MAR peak to be at the position of our superconducting gap, \ie, for $n=2$, $e\vsd=\Delta_\text{Al}$. For a more accurate estimate of $\Delta_\text{Al}$ we perform a linear fit through zero for the six MAR peak positions and find $\Delta_\text{Al}=0.212$~meV which translates to a critical temperature of our Al $\tcal=\Delta_\text{Al}/1.764~k_\text{B}T=1.39\pm0.03$~K~\cite{Tinkham}. This is confirmed by an independent measurement of the $\tc$ of an Al stripline (not shown) and is in good agreement with the critical temperature observed in Fig.~\ref{fig4}.
\paragraph{}
In Fig.~\ref{fig2}b higher-order MAR peaks become increasingly hard to resolve because of the hyperbolic relation of $\vsd$ with $n$. In Fig.~\ref{fig2}c, we therefore convert the x-axis from $\vsd$ to units of $n=2\Delta_\text{Al}/eV_\text{SD}$, resulting in evenly spaced orders of $n$, and plot the conductance for positive bias for the same three $\vbg$ as in Fig.~\ref{fig2}b. We see that for $n>6$, the conductance peaks can no longer be unambiguously assigned and they span multiple $n$. Possibly, the peak patterns for high-order MAR are a superposition of many overlapping MAR processes.
\paragraph{}
Comparing the blue curve in Fig.~\ref{fig2}c with the orange and green curve taken at different $\vbg$, we observe that peak positions for lower order MAR do not perfectly reproduce, for instance, the orange curve does not show a clear peak at $n=4$. We partly explain this by considering our interface transparencies which act as a weak confinement potential, resulting in highly broadened energy levels in the wire. The relative position of these levels with respect to the Fermi level changes the resonant condition for MAR, resulting in a shift of the MAR peak positions~\cite{Buitelaar2003}. Inspecting Fig.~\ref{fig2}a, the high-order MAR peaks ($n>5$) are indeed modulated, both in intensity and position in $\vsd$, by the changing charge and subband population represented by the black regions at low bias ($\vsd<0.05$~mV).
\paragraph{}
We conclude from Fig.~\ref{fig2} that the resolvability of MAR up to $n=6$ means that quasiparticles can elastically scatter at least 6 times on the interfaces, each time traversing the nanowire channel. This requires very low inelastic scattering probabilities and a high (though finite) interface transparency~\cite{Flensberg1988}.

\section{Temperature dependence of $\ic$ and MAR}
We will now investigate the temperature dependence of the switching current and the multiple Andreev reflections. In Fig.~\ref{fig4}a we plot the differential resistance $\dvdi$ versus $\is$ as a function of temperature $T$. The black region $\dvdi=0$ indicates superconductivity and we can see the decrease of $\ic$ for increasing $T$ until $\ic$ disappears at $T\approx1.4$~K, in agreement with $\tcal=1.39$~K calculated from $\Delta_\text{Al}$. \textcolor{black}{The slight increase of $|\ir|$ between $T=0$ - $0.7$ K could be due to changes in the thermal conductivity of the devices’ surroundings, leading to better thermalization at higher temperatures.}

\begin{figure}
  \includegraphics{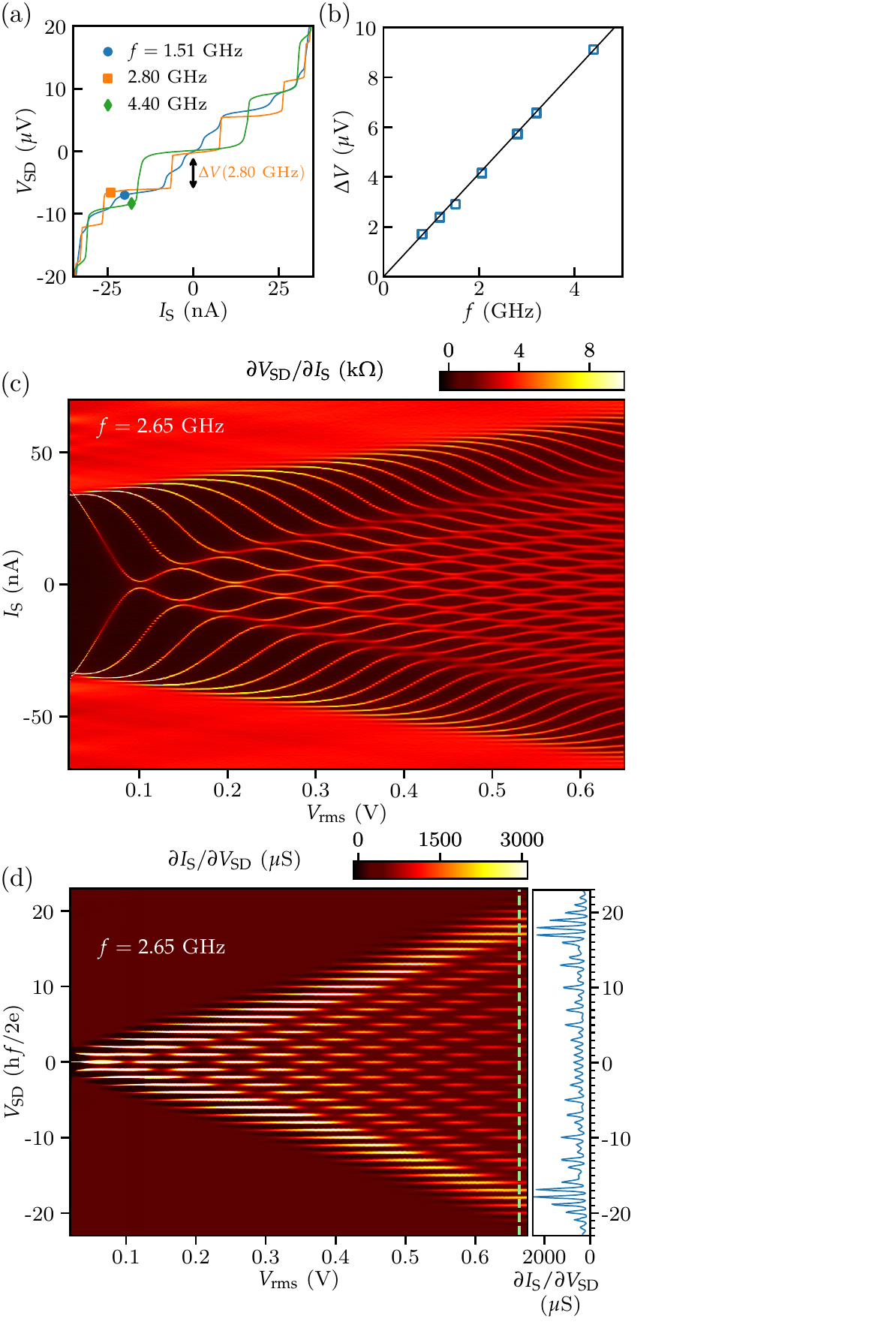}
  \caption{\textbf{ac Josephson effect, up to 23 Shapiro steps.} a) $\vsd$ versus $\is$ for frequencies $f=1.51$, $2.80$ and $4.40$~GHz at respective amplitudes $\vrms=0.11$, $0.13$ and $0.13$~V. $\vrms$ values are ac amplitudes applied before filtering. $f=2.8$~GHz results in a step height of $\Delta V=5.8$~$\mu$V (black arrow). b) Step height $\Delta V$ versus microwave frequency $f$ extracted from data (blue boxes). The black line is a plot of $\Delta V=hf/2e$. c) Differential resistance $\dvdi$ versus $\is$ and microwave rms voltage $\vrms$ applied at a frequency of $2.65$ GHz at $\vbg=-15$ V (red curve in Fig.~\ref{fig1}b). d) Left: $\didv$ versus $\vsd$ and $\vrms$, same measurement data as c) with $\is$ and $\vsd$ axes reversed before numerical derivation (see Methods). $\vsd$ shown in units of $hf/2e$. Right: linecut at $\vrms=0.66$ V showing 23 peaks.}
  \label{fig3}
\end{figure}

\textcolor{black}{
For ballistic supercurrent through a superconductor - normal metal – superconductor Josephson junction, the critical current was modeled by Galaktionov and Zaikin \cite{Galaktionov2001} based on the Eilenberger equations. Note, that this model neglects the spin-orbit interaction, but arbitrary barrier transparencies can be included, and an average is provided over multiple modes. We have used this model to fit our $I_\mathrm{c}(T)$ data. As input parameters we used the critical temperature of 1.4~K, a Fermi velocity of $1 \times 10^5$ m/s, and an electrode separation of 50~nm. These parameters completely determine the shape of the $I_\mathrm{c}(T)$ and provide a zero-temperature estimate for the average critical current of 6~nA per mode. For the experimentally measured $I_\mathrm{c}$, this would correspond to about 7 modes in the junction, consistent with our estimate of the number of modes based on the normal transport data. We furthermore obtain an average transparency of $\sim50$~\%, lower than the previously obtained transparency of $\sim80$~\%~\cite{Ridderbos2017} using the BTK model~\cite{Blonder1982}, averaged over a large gate voltage. This difference could be explained by the fact that $\ic(T)$ was only determined at a single value of $\vbg$ and that $\ic$ is likely to be suppressed with respect to the actual critical current of the junction. The resulting model $I_\mathrm{c}(T)$ has been plotted in Fig.~\ref{fig3}a, together with the experimental data.}
\paragraph{}
The MAR signatures visible outside the superconducting region, scale with $\Delta_\text{Al}$ and therefore gradually decrease and converge to $\vsd=0$ for $T\rightarrow\tcal$. Figure~\ref{fig4}b shows the same dataset as Fig.~\ref{fig4}a converted to a voltage-biased plot (see Methods) and since for MAR order $n=2$, $\vsd=\Delta_\text{Al}$, we have a direct measure of $\Delta_\text{Al}(T)$. We use the BCS interpolation formula~\cite{Senapati2011,Kajimura}:
\begin{equation}\label{bcs}
\Delta (T)%=1.76k_\text{B}T_\text{C,Al}\tanh \left(1.74\sqrt{\frac{T_\text{C,Al}}{T}-1}\right) 
= \frac{2\Delta_{\text{Al},0}}{n}\tanh \left(1.74\sqrt{\frac{\Delta_{\text{Al},0}}{1.76k_{B}T}-1}\right) ;
\end{equation}
where we replaced the pre-factor $\Delta_{\text{Al},0}$ with $2\Delta_{\text{Al},0}/n$ where $\Delta_{\text{Al},0}=0.212$~$\mu$eV is the superconducting gap of Al at $T\approx0$ as determined in Fig.~\ref{fig2}b. We plot this curve in Fig.~\ref{fig4}b and find excellent agreement for the $n=2$ MAR peak and a good fit for $n=3$. The value of $\Delta_{\text{Al},0}$ corresponds to the observed $\tcal\approx1.4$~K, while $\Delta_{\text{Al}}$ follows the BCS curve as a function of T, i.e., the MAR are indeed an excellent measure for the superconducting gap of the Al contacts. 

\section{Shapiro steps}
\label{sec:shapiro}
We now look at the ac Josephson effect by irradiating our junction with a $\lambda/4$ antenna located \textasciitilde$5$~mm above the chip with frequencies ranging from 0.8 to 4.4 GHz. Figure~\ref{fig3}a shows $\vsd$ versus $\is$ for three different frequencies at finite microwave amplitudes $\vrms$, revealing Shapiro steps in the current-voltage relation. Shapiro steps~\cite{Shapiro1964b} are a direct manifestation of the ac Josephson effect where phase locking occurs between the quasiparticles in the junction and the applied microwaves. 
\textcolor{black}{Starting from the ac Josephson relation $V=\frac{\hbar}{2e} \frac{d\phi}{dt}$, quasiparticles can acquire a phase of $\phi=2\pi m$ per period of the applied microwave frequency $f$ with $m$ an integer denoting the Shapiro step number. We can thus write $\frac{d\phi}{dt}=2\pi m f$, translating to a total dc voltage $\vsd=m\Delta V=mhf/2e$~\cite{Tinkham} where $m$ is determined by $\is$ and microwave amplitude $\vrms$. The extracted step height for various frequencies in Fig.~\ref{fig3}b shows good agreement.} We attribute the qualitative variation in the rounding of the steps as a function of frequency to spectral broadening of the microwaves, in turn caused by the microwave antenna properties and the coupling to the Faraday cage in which the sample resides.
\paragraph{}
We now fix the applied frequency at $2.65$~Ghz and plot $\dvdi$ versus $\is$ and $\vrms$ at $\vbg=-15$~V in Fig.~\ref{fig3}c. \textcolor{black}{When increasing $\vrms$, clearly visible lines of differential resistance enter the bias window, each corresponding to a stepwise increase of $\vsd$ by $m\Delta V=$~constant on the plateaus enclosed by the steps.}
\textcolor{black}{As in most experimental setups, our microwave source and antenna have a much higher impedance than our superconducting Josephson junction~\cite{Tinkham,Gross2005} and it therefore acts as an ac current source. Therefore, the width of the current plateaus cannot be described by simple Bessel functions, but can only be numerically approximated~\cite{Gross2005,Russer1972,Tinkham}.}
\paragraph{}
\textcolor{black}{To gain insight in the number of Shapiro steps and their corresponding plateau heights in Fig.~\ref{fig3}c, we show a $\vsd$ biased plot in units of $\Delta V=hf/2e$ of the same measurement data in Fig.~\ref{fig3}d. The plateaus of constant $\vsd$ in Fig.~\ref{fig3}c are now visible as peaks in differential conductance $\didv$.} Looking at the linecut on the right we see that up to 23 steps are visible, all aligned with values of $m\Delta V=mhf/2e$. This clear demonstration of the ac Josephson effect in Fig.~\ref{fig3}, together with dc effects such as MAR and finite a $\ic$ is proof that our junction is, indeed, a well behaved Josephson junction.

\begin{figure*}
  \includegraphics{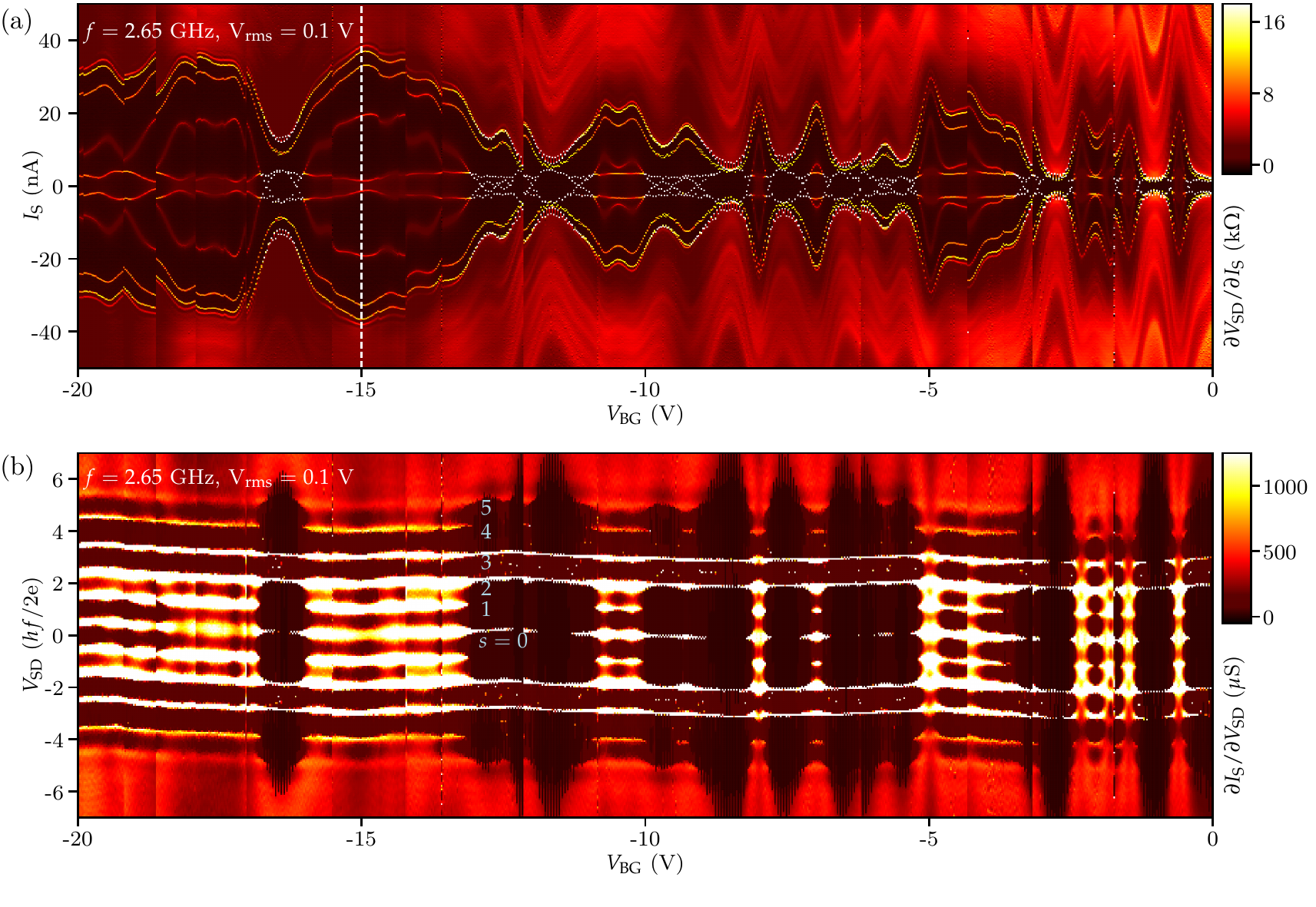}
  \caption{\textbf{Shapiro steps versus backgate voltage.} a) Differential resistance $\dvdi$ versus $\is$ and $\vbg$ at fixed microwave frequency $f=2.65$ GHz and $V_\text{rms}=0.1$~V. The direction in which $\is$ is swept is reversed after each step of $V_\text{BG,step}=25$~mV. The white dashed line denotes $\vbg=-15$~V used in Fig.~\ref{fig3}. b) Differential conductance $\didv$ vs $\vsd$ (units of $hf/2e$) vs $\vbg$. Same data set as a) where $\is$ and $\vsd$ axes were inverted (see section Methods). We identify up to 5 Shapiro steps, \ie, current plateaus $s=0\text{ - }5$. The `wavy' pattern, \ie, the gradual shift of the lines in $\vsd$ as a function of $\vbg$, is caused by small variations in the leakage current at different $\vbg$.}
  \label{figA8_1}
\end{figure*}

\section{Shapiro steps versus $\vbg$}
Previously, $\vbg$ was fixed at $-15$~V, corresponding to a region with a high $\ic$ and low hysteresis, i.e. close to critical damping corresponding to a Stewart-McCumber parameter $\beta_{C}$ close to 1~\cite{Tinkham}. In Fig.~\ref{figA8_1}a we show a current-sourced backgate dependence of Shapiro steps at fixed microwave frequency and power. The junction is generally hysteretic for regions with lower $\ic$, observed in this figure at regions where the Shapiro steps are moving closer together on the $\is$ axis. This corresponds to a higher normal state resistance $R_{N}$ which increases $\beta_{C}$ and results in an underdamped junction. Since measurement data were acquired in both directions while sweeping $\is$ back and forth (after stepping $\vbg$ with $25$~mV after each sweep), hysteresis appears as a white speckle pattern caused by the data acquisition alternating directions in $\is$ (see for instance between $\vbg=-10$ and $-8$~V). The observed oscillations of $\ic$ (and indirectly $R_{N}$) as a function of $\vbg$ are again the result of the varying population of subbands.
\paragraph{}
Figure~\ref{figA8_1}b shows the voltage-biased backgate dependence of the same measurement data as Fig.~\ref{figA8_1}a where plateaus in current are translated to peaks in $\didv$ by inverting the $\is$ and $\vsd$ axis and normalising $\vsd$ to $\Delta V$. We identify 5 Shapiro steps which partially disappear in regions with increased $R_{N}$ when the junction is hysteretic. For this specific $\vrms$, steps 0, 1, and 4 disappear in the hysteretic regions, loosely corresponding to the smaller current plateau widths. Since the plateau widths vary (Bessel-like) with $\vrms$ (see Fig.~\ref{fig3}c), which steps are missing therefore changes as a function of $\vrms$ (not shown here). At $\vbg=-15$~V, all steps are visible which is the reason this specific voltage was used in Fig.~\ref{fig3}.

\section{Conclusion}
We have realized a Josephson junction where the high interface transparency between the superconducting leads and the nanowire results in multiple Andreev reflections up to the 6\textsuperscript{th} order. We additionally show up to 23 Shapiro steps, clearly demonstrating the ac Josephson effect for the first time in this system. 
\textcolor{black}{We estimate the total contact transparency to be between 50\% and 80\% based on the temperature dependence and previously obtained results. We furthermore estimate the nanowire segment to be in the ballistic limit and improving the contact interfaces could therefore result in fully ballistic junctions.}
\paragraph{}
Ge-Si nanowire-based Josephson junctions possess all ingredients necessary for obtaining Majorana fermions and in parallel experiments we have found very hard induced superconducting gaps~\cite{Ridderbos2017a}. We therefore propose a follow-up experiment with a device design suitable for probing the zero-energy Majorana bound states in the nanowire~\cite{Mourik2012b}. \textcolor{black}{Additionally, other applications such as superconducting qubits and Andreev (spin) qubits, can now actively be pursued in this system.}

\section{Methods}
\subsection{Post-processing of measurement data}\label{ivconversion}
All measurements in this work are performed using a 3-probe measurement. A series resistance of $3.46$~k$\Omega$ was subtracted from all measurement data. In Fig.~\ref{fig2}, Fig.~\ref{fig4}b, Fig.~\ref{fig3}d and Fig.~\ref{figA8_1}b, the datasets are obtained using a current source driving $\is$ with $\vsd$ measurement after which a software routine is used to invert the source and measurement axis. To obtain equidistant points on the new $\vsd$ source axis, the points are recalculated by interpolation in $\is$ on a grid with predetermined $\vsd$ stepsize. The resolution of $\vsd$ is chosen high enough so that no features in the original measurement of $\vsd$ are lost. In Fig.~\ref{fig3}c and Fig.~\ref{fig3}d a similar grid interpolation procedure was used to convert the $x$-axis from units of dBm to V.

\subsection{MAR dataset acquisition}\label{mardata}
In the open regime MAR peaks can only be seen when the junction is in the dissipative current state and since higher-order ($n>6$) MAR reside close to $0$ bias, they can be obscured by the superconducting 'blind spot' of the junction. We use the bi-stable current-voltage relation (hysteresis) of our underdamped Josephson junction in Fig.~\ref{fig2}a, where we choose a region of $\vbg$ with a low $\ir$. In order to measure the current-voltage relation of the junction way below $\ic$, the low $\ir$ is exploited by sweeping from finite $\abs{\is}$ to $0$ in both bias directions. We note that $\ir$ is still finite which is reflected in the small black oscillating `blind spot' region around $\abs{\vsd}=0$, although a much larger range of $\vsd$ can now be probed. The visibility of MAR also depends on $\gn$: a higher $\gn$ results in a lower voltage drop over the same $\is$, thus effectively enhancing measurement resolution of the equipotential MAR peaks. This is especially important for higher order MAR ($>4$), since its hyperbolic relation with $\vsd$ means that the corresponding peaks become very closely spaced.

%\paragraph{}
%\section{Acknowledgements}
\section{Acknowledgements}
\begin{acknowledgments}
%\begin{acknowledgement}
F.A.Z. acknowledges financial support through the Netherlands Organization for Scientific Research (NWO). E.P.A.M.B. acknowledges financial support through the EC Seventh Framework Programme (FP7-ICT) initiative under Project SiSpin No. 323841.
\end{acknowledgments}
%\end{acknowledgement}
%\bibliography{library}

%Control: production of eprint (-1) disabled
%

\end{document}